\documentclass[aps,pre,onecolumn,showpacs,a4paper]{revtex4}
\usepackage{graphicx} 
\usepackage{amsmath}
\usepackage{amssymb}
\usepackage{amscd}
\usepackage{soul}
\usepackage{cancel}
\newcommand{\thickone}{\mbox{$1\!\!1$}}

\begin{document}

 \title{Fluctuation Relations for Quantum  Markovian Dynamical Systems}
\author{R. Chetrite}
 \affiliation{Physics of Complex Systems, Weizmann Institute of Science,
  Rehovot 76100, Israel}
\author{K. Mallick}
 \affiliation{Institut de Physique Th\'eorique, CEA Saclay, 91191
  Gif-sur-Yvette, France}
 \date{February 4, 2010.}
 
\begin{abstract}  
     We derive a  general  set of fluctuation relations  for
 a  nonequilibrium  open quantum  system  
 described by a Lindblad  master equation.  In the special case
 of  conservative   Hamiltonian  dynamics, these  identities allow
 us to retrieve  quantum versions  
  of Jarzynski and Crooks relations. 
 In the linear response regime,  these  fluctuation relations yield a  
 fluctuation-dissipation theorem  (FDT) valid for a stationary state
 arbitrarily far from equilibrium. For a closed system, this FDT reduces to 
 the  celebrated  Callen-Welton-Kubo formula. 
\end{abstract} 
 \maketitle 

           Fluctuations in nonequilibrium systems have been shown to
  satisfy various remarquable  relations 
  \cite{GallavottiCohen,Jarzynski} discovered  during  the last fifteen years.
  These results  have led to fierce  discussions  concerning
 the nature of heat, work and entropy, raising 
 the  fundamental issue  of
  understanding  the interactions between a given system and its environment
 (e.g., a thermal bath). In the  classical realm, these  problems 
  have been progressively clarified while they are still under
  investigation in the quantum world. Many  works 
 on quantum fluctuation relations, including
 the  pioneering  ones,   consider  only   closed systems 
 prepared in a Gibbs state  and  isolated from their
 environment during their evolution which  is thus   unitary
 \cite{Kurchan,Tasaki,Talkner1,Gaspard1,SMukamel}.  
 The general case  of an  open system  continuously interacting 
 with its surroundings  can be formally reduced to the previous
 situation   by considering  the system together
 with  its environment  to be  a  closed global  system:
 this is the approach adopted in \cite{Kurchan,Jar,Gaspard2,Campisi} 
 (see \cite{SMukamel} for a review).
  However, one must   project out 
 the degrees of freedom of the bath  to derive an effective dynamics
  for the system of interest. It is known in classical mechanics,
 that after such an elimination procedure
 the Hamiltonian dynamics for the global system  becomes 
  effectively a stochastic dynamics for the initial  system.  Similarly, 
 in the quantum case, after integrating out the degrees 
 of freedom of the   environment and using a  Markovian approximation, 
  a quantum
 master equation for the density matrix of the  system is obtained.
 Under some  further assumptions \cite{Lindblad,Gardiner},
   this master equation can be 
 brought into a  form  known as the  Lindblad equation. 
 The Lindbladian evolution is   a non-unitary dynamics for the
 density matrix  $\rho_t$ of the open  system, described by a 
 differential equation  with  generator $L_t$.
 This  effective Markovian description  
   is  widely used in Quantum Optics \cite{Gardiner}.

  In the present work, we  study
  the time-reversal properties of 
  an  open quantum system  modeled by a
  Lindblad  equation. A similar point of view was adopted in previous works; 
   in  \cite{Yukawa,Maes,CrooksQM},
  the time evolution was discretized in an  
 ad-hoc manner;  in \cite{SMukamel0,SMukamel2,Roeck}
 the quantum master equation was treated as  an
  effective classical master equation and 
 the concept  of trajectory and  the  fluctuation relations
  for classical system were  used. 
 Our approach, here, is to work directly with 
 the continuous time Lindblald equation without
  referring to any  classical effective system, 
 to define an associated time-reversed dynamics and to derive 
 fluctuation relations 
 with  quantum observables.

 The key  results of this  work are  given
  in Eqs.~(\ref{fluctuat},\ref{QuantFT},\ref{QuantJarz},\ref{TFDG}) and 
  represent an original contribution  to  quantum  
   non-equilibrium statistical mechanics. 
     Thanks to a  suitable deformation
 of the master equation, we   prove  a generic relation 
 amongst correlation functions,  a kind
  of book-keeping formula 
  which    yields   the quantum analog of Jarzynski  and Crooks relations.  
  Furthermore,  by  a  lowest order expansion,  we derive 
 a  generalized fluctuation-dissipation theorem  valid
 in the vicinity of  a quantum  non-equilibrium steady state.  
 For the special case of a closed system, our approach retrieves
 previously known work identities
 \cite{Kurchan,Tasaki,Talkner1,Gaspard1,Yukawa,Maes} as well as  
 the quantum equilibrium   fluctuation-dissipation theorem
 \cite{Callen,Kubo}.

    The density matrix $\rho_t$ of the 
   open quantum  system  prepared initially
   with  $\rho_0$,  evolves according to the master
 equation
 \begin{equation}
 \partial _{t}\rho _{t} = L_{t}^{\dagger} \, \rho _{t}
\label{eq:master1}
 \end{equation} 
 where $L_{t}$ is the Lindbladian superoperator
(i.e. a linear map in the space of operators) and  $L_{t}^{\dagger}$  its
 adjoint with respect to the   operator scalar product 
 $(X,Y) = Tr(X^{\dagger} Y)$, $X$ and $Y$
being  arbitrary  operators  and $X^{\dagger}$  the hermitian conjugate of $X$.
Under suitable hypotheses (such as trace preserving and complete  positivity 
 \cite{Lindblad,Gardiner})  the  
 Lindbladian  $L_t$  takes  the  generic form: 
 \begin{equation}
 L_t \,  X =i[H_t,X] -\frac{1}{2}\sum_{i=1}^{I}
\left( A_{t}^{i \dagger }A_{t}^{i}X + XA_{t}^{i \dagger}A_{t}^{i}
  -2A_{t}^{i \dagger }XA_{t}^{i}\right)  \,. 
\label{def:Lindbladian}
 \end{equation}   
 The first term $i[H_t,X]$ is the conservative part,
   $H_{t}$ being the Hamiltonian
 of the system that may depend on time. The second term models the interactions
 with the environment  (dissipation and coherence effects):  the $A_{t}^{i}$'s
 are, in general, non-hermitian  operators  that may  depend explicitly 
  on time. In a closed  system, 
  $A_{t}^{i} \equiv 0.$
 To the time-dependent  Lindbladian  $L_{t}$ we associate
 the  {\it accompanying} \cite{Hanggi}
  density-matrix $\pi_t$ such that  
  $L_{t}^{\dagger} \, \pi_t = 0$. Physically,  $\pi_t$  represents
 the stationary  density-matrix in a  system where time is  frozen
 at  its instantaneous value $t$. However, because $\pi_t$ depends 
 on time, it does not satisfy Eq.~(\ref{eq:master1}). For a closed
 system, we have  $\pi _{t}=Z_{t}^{-1}\exp (-\beta H_{t}).$

  The formal
 solution of Eq.~(\ref{eq:master1})   can be written 
  as  $\rho _{t} =  \left(P_{0}^{t} \right)^{\dagger} \rho _{0}$ 
 where the evolution  superoperator is  given by 
\begin{eqnarray}
 P_{s}^{t}  &=& \overrightarrow{\exp }\left( \int_{s}^{t} L_{u} \,du\right)  \equiv
\sum_{n}\int_{s\leq s_{1}\leq s_{2}\leq\ldots\leq s_{n}\leq t}
 L_{s_{1}} L_{s_{2}}\ldots L_{s_{n}} \,\prod_{i=1}^n ds_{i} .
\label{eq:ExpOrd1}
 \end{eqnarray} 
 For $0\leq t_{1}\leq t_{2}\leq \ldots \leq t_{N}\leq T$, the
  time-ordered   correlation  of observables  $O_{0},O_{1},O_{2}...O_{N}$ is defined as \cite{Gardiner}
\begin{eqnarray}
&\left\langle O_{1}(t_{1})O_{2}(t_{2}) \ldots O_{N}(t_{N})\right\rangle 
&=  Tr\left( \pi_{0}P_{0}^{t_{1}}O_{1}P_{t_{1}}^{t_{2}}O_{2} \ldots 
P_{t_{N-1}}^{t_{N}}O_{N}\right) \, .
\label{def:Correlations}
\end{eqnarray}
 Note that  $P_{t_{i}}^{t_{i+1}}$ operates on all the terms to its right
 and that  the initial density matrix is given by $\rho_0 = \pi_0$.

 A crucial element  in our approach is the 
 {\it  time-reversed} system,  characterized
 by the following  Lindbladian: 
  \begin{equation}
   L_{t^*}^R=   K \pi_t^{-1} L_{t}^{\dagger} \pi_t K \,\,\,\, 
 \hbox{ with } t^* = T -t \, .
\label{def:LindbReversed}
\end{equation}
The  superoperator  $K$  acts on an operator $X$
 as  $K \, X = \theta X \theta^{-1}$  \cite{Agarwal}, 
    $\theta$ being  the  time inversion  
 anti-unitary operator, with  $\theta^2=1$, that 
 implements  time-reversal on the states $\psi$ of 
 the Hilbert space.  The  superoperator  $K$
 is antiunitary, with  $K^2=1$,  $K=K^{-1}=K^{\dagger}$, and is multiplicative
 i.e.  $K(XY)=K(X)K(Y)$.  Note  that in  Eq.~(\ref{def:LindbReversed}), 
  $\pi_t$ and  $\pi_t^{-1}$ denote  left-multiplication superoperators.
   It is a non-trivial  fact  that  the r.h.s of Eq.~(\ref{def:LindbReversed}) 
 defines  a {\it bona fide}
   Lindbladian. This  is ensured, for example,  by
  imposing  at each time $t$ the quantum  microreversibility condition (or detailed balance)
  \cite{Agarwal},  which implies  $L_{t^*}^R = L_{t}$. Here, we do  not assume 
 detailed balance and we  only require the weaker  condition that  $L_{t^*}^R$ is a 
 well-defined  Lindbladian.
  Using  Eq.~(\ref{def:LindbReversed})  and the relation 
   $L_{t} \, 1 = 0$,  we find
 $(L_{t^*}^R)^{\dagger} K \, \pi_t = 0$ and hence 
  $\pi_{t^*}^R =  K \, \pi_t$,
 thus relating the   accompanying distribution of  the time-reversed system
 with that   of the original system.

   From   Eqs.~(\ref{eq:ExpOrd1}) and (\ref{def:Correlations}),  the  evolution
  superoperator  of the   time-reversed system  is  $ P_{s}^{t,R}=
 \overrightarrow{\exp }\left( \int_{s}^{t}  L_{u}^R \, du \right) $
 and  the  time-ordered  correlations   are:
\begin{eqnarray} 
\left\langle O_{1}(t_{1})O_{2}(t_{2})\ldots O_{N}(t_{N})\right\rangle^R  
= Tr\left( \pi_{0}^RP_{0}^{t_{1},R}O_{1}P_{t_{1}}^{t_{2},R}O_{2} \ldots 
P_{t_{N-1}}^{t_{N},R}O_{N}\right) \, .
\label{eq:CorrelRenv}
 \end{eqnarray} 

  Given a scalar $\alpha$,
 we deform the Lindbladian $L_{t}$ and $L_{t}^R$ 
 to superoperators $L_{t}(\alpha)$ and $L_{t}^R(\alpha),$ 
 that act  on an observable  $X$  as follows:
\begin{eqnarray}
L_{t}(\alpha) \,  X &=&   \left( L_{t}  
  + \alpha \pi _{t}^{-1}\partial _{t}\pi _{t} \right) X   \nonumber \\
  \hbox{  and } \,\,\,
L_{t}^R(\alpha) \, X &=& \left( L_{t}^R  
  + \alpha (\pi_t^R)^{-1}\partial _{t}\pi_{t}^R \right) X   \,.  
\label{def:LindbDef}
\end{eqnarray}
The corresponding  evolution superoperators are 
    $ P_{s}^{t}(\alpha)=
 \overrightarrow{\exp }\left( \int_{s}^{t}   L_{u}(\alpha)\, du \right) $ and 
 $ P_{s}^{t,R}(\alpha)=
 \overrightarrow{\exp }\left( \int_{s}^{t}   L_{u}^R(\alpha) \, du \right) \, . $
  Furthermore,  the following  conjugation identity
 between  superoperators is  satisfied:
 \begin{equation}
  \pi_0 P_{0}^{T}(\alpha) = 
 \left[    \pi_T  K  P_{0}^{T,R}(1- \alpha) K \right]^{\dagger}  \, .
\label{eq:Conjug1}
\end{equation}
 This relation 
 stems from the fact  that the operator  
$U_t =  \pi_0 P_{0}^{t}(\alpha)  \pi_t^{-1}$ satisfies the evolution
 equation 
 $\partial_t U_t = U_t \left(K L_{t^*}^R (1- \alpha)  K \right)^{\dagger} $. 
 Equation~(\ref{eq:Conjug1})  is the key duality relation,
 similar to the   identity  that lies  
  at the heart of the proof of the classical
 Jarzynski identity \cite{Jarzynski}  and of the Gallavotti-Cohen
 theorem  in Langevin and Markovian systems \cite{Kurchan2}. 
 Besides,  all these derivations rely on   a suitably 
  modified  dynamics with respect to a continuous parameter.

 Applying the fundamental identity~(\ref{eq:Conjug1})
  to two arbitrary observables 
 $A$ and $B$ leads to (using  the fact that $K$ is multiplicative  and anti-unitary)
\begin{equation}
Tr\left( B^{\dagger}\pi _{0}P_{0}^{T}(\alpha) A \right) = Tr\left(
\left(K \, A^{\dagger}\right) \pi_{0}^R P_{0}^{T,R}(1-\alpha )  \left( K \, B \right) \right)  \, .
\label{fluctuat}
\end{equation}
 This equation, which  is  the   essence of  the quantum
 fluctuation theorem, expresses a  generalized detailed balance condition. 
   For  a stationary state $\pi_t = \pi$ 
 (so that the  $\alpha$-dependence drops out)
 that is also  reversible (i.e. $P_{0}^{T,R}=P_{0}^{T}$), Eq.~(\ref{fluctuat})
 becomes equivalent to the detailed balance condition of \cite{Agarwal}.

  Equation~(\ref{fluctuat})   can be brought into a more familiar form 
 by introducing
\begin{equation}
   W_t = -(\pi_t)^{-1} \partial_t \pi_t \,\,\, \hbox{  and } \,\,\, 
   W_t^R = -(\pi_t^R)^{-1} \partial_t \pi_t^R \, ,
  \label{def:work}
 \end{equation}
 (in the classical limit, these  operators   reduce to the injected  power), 
and by using the following relation, valid for two operators $X$ and $Y$
\begin{equation}
 Tr\left( \pi_{0} Y  P_{0}^{T}(\alpha) \, X \right) 
 = \left\langle Y(0)  \,\,  \overrightarrow{\exp }
 \left( -\alpha \int_{0}^{T}  W_{u} \, du\right)  X(T) \right\rangle \,.
\label{def:Dyson}
\end{equation}
 This   formula is proved  by expanding  $P_{0}^{T}(\alpha)$
 w.r.t. the deformation parameter $\alpha$ (Dyson expansion), 
  rewriting the trace as
 a correlation function via 
 Eq.~(\ref{def:Correlations}) and finally identifying the result with the r.h.s.
  \footnote{In full rigor, 
 the integrand is   $\int_{0}^{T}du \,  W_{u}(u)$
 because   a  supplementary time 
 dependence is introduced   through the use of 
 Eq.~(\ref{def:Correlations}),   that should appear as the argument of the time-dependent $W_u$ operator.}.
   Inserting the
 symmetry  relation~(\ref{eq:Conjug1}) into equation~(\ref{def:Dyson}),
 the  Fluctuation Relation for an open quantum  Markovian
 system is obtained: 
\begin{eqnarray}
&\left\langle \left( \pi _{0}B\pi _{0}^{-1}\right)^{\dagger}\left( 0 \right)
\overrightarrow{\exp }\left( -\alpha \int_{0}^{T} W_{u} \, du\right)
A(T)\right\rangle  =  \label{QuantFT} \\  & \left\langle \left( \pi _{0}^{R} \left( K \, A \right)
\left( \pi_{0}^{R}\right) ^{-1}\right)^{\dagger}\left( 0 \right)
\overrightarrow{\exp }\left( -(1- \alpha )\int_{0}^{T} W_{u}^{R} \, du\right)
 \left( K \, B \right)(T)\right\rangle ^{R}  \, .  \nonumber 
\end{eqnarray}
  This   identity  is original,  it encodes  the main results of our  work
  and will allow  us to derive   various relations 
 for quantum systems far from equilibrium.
 If we interpret the mean values as classical
 averages and the operators as commuting c-numbers,
 then Eq.~(\ref{QuantFT}) becomes
 Crooks' relation.  In the quantum case, if we take
 $A = B = \thickone$ and $\alpha =1$, 
 then  Eq.~(\ref{QuantFT}) yields a quantum analog
 of  the Jarzynski identity:
\begin{equation}
  \left\langle \overrightarrow{\exp }
 \left( -\alpha \int_{0}^{T} W_{u} \, du \right) \right\rangle  = 1  \, . 
\label{QuantJarz}
\end{equation}
 This quantum  Jarzynski relation 
 was first derived  in  \cite{Yukawa}, but   the  operator-ordering  issue
 was not accurately 
 taken into account \footnote{Note that  the 
   Jarzynski relation does not involve the time-reversed dynamics. Thus,
 the proof of Eq.~(\ref{QuantJarz}) does not require
 to assume that $L_{t^*}^R$ is a  Lindbladian. In particular,
 no  quantum  microreversibility condition is needed.}.

 We now   derive a Quantum Fluctuation-Dissipation
 Theorem for a  system in the vicinity of a non-equilibrium steady state.
 Suppose that the Lindbladian is given by
 $ L_{t}= L_0- h^{a}(t) M_{a} $  
 where  $L_0$ is time-independent with invariant density-matrix
 given by $\pi_0$.
 The time-dependent
 perturbations  $h^{a}(t)$   are supposed to be small
 and a summation over the repeated index $a$ is understood.
 At first order, the accompanying  density-matrix $\pi_t$,
 with  $ L_{t}^{\dagger} \, \pi_t =0$,  is given by
$\pi_t = \pi_0 + h^{a}(t)  \epsilon_a$ where $\epsilon_a$ satisfies
 $ L_{0}^{\dagger} \, \epsilon_a =M_{a}^{\dagger} \, \pi_0$ and 
   $W_t$ defined in~(\ref{def:work}) reads
 $ W_{t}\equiv -\dot{h}^{a}(t)  D_{a}  $  with 
 $  D_{a} =  \pi_0^{-1}  \epsilon_a   $. 
  Using  Eq.~(\ref{QuantFT}) with an 
 arbitrary operator $A$, with  $\alpha =1$ and 
 $B= \thickone$, and taking
 its  functional derivative w.r.t. $h^a(u)$ with $u<T$, we obtain 
\begin{equation}
 \left. \frac{\delta \left\langle A\left( T \right)\right\rangle }{\delta h^{a}(u)}
\right| _{h=0} +  \left. \frac{\delta \left\langle \overrightarrow{\exp }\left(
-\int_{0}^{T}W_{v} \, dv \right) A\left( T \right)\right\rangle _{0}}{\delta h^{a}(u)}
\right| _{h=0}  = 0 \, .
  \end{equation}
   Using  the  first order expansions derived above, we obtain 
\begin{equation}
\left. \frac{\delta \left\langle A\left( T \right)\right\rangle }{\delta h^{a}(u)}
\right| _{h=0}=\frac{d}{du}\left\langle D_{a}(u)A\left( T \right)\right\rangle _{0} \, .
\label{TFDG}
\end{equation}
 We emphasize that the expectation value on the r.h.s. is taken
 with respect to the unperturbed density matrix $\pi_0$.
 By choosing  $A_{T}=D_{b}(T)$,
 Eq.~(\ref{TFDG})   becomes structurally similar  to 
 the usual equilibrium fluctuation dissipation theorem. This  generalizes  to the quantum case,
  a     recently
 obtained result for classical systems \cite{Prost} (see  \cite{Weidlich} for  an alternative approach).

   In the last part of this work, we consider the special case of
 an isolated system. Here, the Lindbladian reduces to the Liouville
 operator: $L_t \,  X =i[H_t,X]$.  The Hamiltonian is time-dependent
 but the evolution of the system is unitary and in this framework
 the general relation~(\ref{QuantFT}) reduces to the recently obtained
 Quantum work relations \cite{Kurchan,Tasaki,Talkner1,Gaspard1}.
  For a closed system,
 the evolution superoperator  on an observable  
 $X$ reduces to the  unitary action
$P_{0}^{T} \, X =\left( U_{0}^{T}\right)^{\dagger}XU_{0}^{T}$, where the evolution operator 
 $ U_{0}^{T}=\overleftarrow{\exp }\int_{0}^{T} \left( -iH_{u} \right)\, du$.
  It can be shown then that  the time-reversed
 system~(\ref{def:LindbReversed}) is also closed, with 
  Hamiltonian $H_{t^{*}}^{R}= K \, H_{t} $  and  evolution
 operator $U_{0}^{T,R} = K \, \left(U_{0}^{T}\right)^{\dagger}$
 \footnote{  For a spin-0 particle in a magnetic field, the time-inversion
 operation~(\ref{def:LindbReversed}) is supplemented by the
 requirement  that the reversed system evolves with  vector potential
  $A_{t^{*}}^{R}= - A_{t}.$}.
    
    Using  identity~(\ref{eq:Conjug1}) for $\alpha=1$  and   the fact that $K$ is multiplicative, we 
 find  that  $P_{0}^{T}(\alpha=1)A$ is equal to 
 \begin{eqnarray}
\pi _{0}^{-1}K \left(\left( P_{0}^{T,R}\right) ^{\dagger }K \left(\pi
_{T} A \right) \right)=\pi _{0}^{-1}K \left( U_{0}^{T,R}K\left( \pi _{T} A \right) \left(
U_{0}^{T,R}\right) ^{\dagger } \right)=   \label{evolfermedef} \\\pi _{0}^{-1}K\left( K\left( \left(
U_{0}^{T}\right) ^{\dagger }\right) K\left( \pi _{T} A \right) K\left(
U_{0}^{T}\right) \right) =\pi _{0}^{-1}\left( U_{0}^{T}\right) ^{\dagger
}\pi _{T} A U_{0}^{T}.
\nonumber
\end{eqnarray}
 Substituting the last  expression   in Eq.~(\ref{fluctuat}) leads to 
\begin{equation}
Tr\left( B^{\dagger}\pi _{0}\pi _{0}^{-1}U_{0}^{T\dagger}
\pi_{T}AU_{0}^{T}\right) = Tr\left( K\left(A^{\dagger }\right)
\pi _{0}^{R}\left(U_{0}^{T,R}\right)^{\dagger}K\left(B\right)U_{0}^{T,R}\right) \, .
  \label{QFTisole}
\end{equation}
 Recalling that 
  $\pi _{0}^{-1}$ and $\pi_{T}$ are given by the 
 Boltzmann law,  the above equation  becomes
 in  the Heisenberg representation denoted by the
 superscript ${\mathcal H}$, 
 \begin{eqnarray}
 Tr\left( B^{\dagger}\pi _{0}\exp (\beta H_{0}^{{\mathcal H} }(0))
\exp \left( -\beta H_{T}^{{\mathcal H} }(T)\right)
 A^{{\mathcal H}}(T)\right)   
  = \frac{Z_{T}}{Z_{0}}
 Tr\left( K\left(A^{\dagger }\right)
\pi _{0}^{R}\left(U_{0}^{T,R}\right)^{\dagger}K\left(B\right)U_{0}^{T,R}\right) \,.
\label{QFTisole2}
\end{eqnarray}
 We emphasize  that for  $B=\thickone$, Eq.~(\ref{QFTisole}) 
 is a tautology (because $K$ is anti-unitary), however
 it implies the non-trivial result~(\ref{QFTisole2}): this 
   feature is characteristic of most of the  derivations of the
 work identities. If we now take   $A=B=\thickone$, we end up
 with the quantum Jarzynski relation for closed systems
 as first found  by Kurchan and Tasaki \cite{Kurchan,Tasaki} :
 \begin{eqnarray}
 Tr\left( \pi _{0}\exp (\beta H_{0}^{{\mathcal H}}(0))
 \exp \left( -\beta H_{T}^{{\mathcal H} }(T)\right) \right)
  =\frac{Z_{T}}{Z_{0}} \,.
\label{QJarz}
\end{eqnarray}
Rewriting the l.h.s.  as a time-ordered exponential and using 
$\frac{d}{dt}H_{t}^{{\mathcal H} }=  (\partial _{t}H_{t})^{{\mathcal H} }$
 we conclude  as  in  \cite{Talkner1}   that 
 \begin{equation}
Tr\left( \pi _{0}\overrightarrow{\exp }
\left( -\beta \int_{0}^{T}\left(
\partial _{s}H_{s}\right) ^{{\mathcal H}}(s)\, ds \right) \right) 
=\frac{Z_{T}}{Z_{0}} \, .
\end{equation}
 Considering now  a system perturbed near equilibrium
 with $H_{t}=H-h_{t}^{a}(t)O_{a}$, we 
 can calculate explicitly the first order perturbation
 to the canonical  density-matrix, 
$\exp (-\beta H_{t})=\exp (-\beta H)+h_{t}^{a}\int_{0}^{\beta } \exp
(-\alpha H)O_{a}\exp (\alpha H)\exp (-\beta H)\,d\alpha$ and find
that 
$
D_{a}=-\beta \left\langle O_a\right\rangle _{0}+\int_{0}^{\beta } \exp
(\alpha H)O_{a}\exp (-\alpha H)\,d\alpha
$. Finally, the Fluctuation-Dissipation Theorem~(\ref{TFDG}) becomes
\begin{eqnarray}
 \left. \frac{\delta \left\langle A\left( T \right)\right\rangle }
 {\delta h^{a}(u)}
 \right| _{h=0} = i\left\langle O_{a} \left(A \left( T-u+i\beta \right) - 
 A \left( T-u \right)\right)\right\rangle _{0} \, .
\label{fdtf} 
\end{eqnarray} 
 To obtain this result,  we performed the  derivation w.r.t. time $u$
  on the r.h.s. of Eq.~(\ref{TFDG}) and then   used 
 an analytic continuation  to write the 
 inverse temperature $\beta$ as  the imaginary part of the  time, as allowed 
 by  the KMS condition \cite{Martin}. 
 This  is the real space version of the
 celebrated  result derived 
  amongst others by Callen and Welton, and by Kubo \cite{Callen,Kubo}.
  For a closed system, an   alternative
   proof that  relation (\ref{QJarz}) implies Eq.~(\ref{fdtf})
 is given in  \cite{Gaspard1}.

  In this work, we have derived  fluctuation relations for an open quantum
 system   described by a  Lindblad dynamics
 that  takes into account the 
 interactions  with the  environment as well as  measurement processes. 
  We prove  the  fluctuations relations thanks to  a suitable deformation
 of the system's dynamics: this  key technical idea 
  provides  a truly  unified picture of the  fluctuations relations,
 whether classical or quantum,  and does not require  to define 
 the concept of work at the quantum level.
 A possible extension of our work would be to  study  
 particles with non-zero spins such as Dirac spinors. 
 Besides,  choosing a  time inversion different from that of   Eq.~(\ref{def:LindbReversed}) 
  may lead to various families of fluctuation relations, 
 as happened  in the classical case  \cite{Chetrite}.
 Other  extensions   would be to derive exact  solutions
 for  specific  models (such as  quantum Brownian motions)  thus providing
  experimentally testable predictions.
 Finally, the investigation  of  time-reversal properties of general
 non-Markovian quantum systems, in which the characteristic 
 time scale of the environment cannot be neglected
 w.r.t. that of the system,  
  should  also yield  interesting  fluctuation relations.

 R. C. thanks  K. Gaw\c{e}dzki for pointing out the fact that 
 the Lindbladian character of Eq.~(\ref{def:LindbReversed}) is non-trivial
 and the relation with detailed balance.
  R. C.  acknowledges the support of the Koshland center for basic research.
 K.M. thanks M. Bauer, S. Mallick and  H. Orland for useful comments.
  Results similar to those presented here
 were also reached  independently by  K. Gaw\c{e}dzki
  and S.  Attal some time ago \cite{Attal}.


\begin{thebibliography}{article}
 \bibitem{GallavottiCohen} D. J. Evans,  E. G. D.  Cohen, G. P. Morriss, 
  Phys. Rev. Lett. {\bf 71},   2401 (1993);
  G. Gallavotti and  E. G. D.  Cohen, Phys. Rev. Lett.  {\bf 74}, 2694 (1995).

\bibitem{Jarzynski} C.~Jarzynski,  Phys. Rev. Lett. {\bf 78}, 2690 (1997);
 C. Jarzynski, Phys. Rev. E {\bf 56}, 5018 (1997); 
 G. E. Crooks, J. Stat. Phys. {\bf 90}, 1481 (1999).
 
\bibitem{Kurchan}  J. Kurchan, arXiv:cond-mat/0007360 (2000).
\bibitem{Tasaki} H.  Tasaki,  arXiv:cond-mat/0009244 (2000).

\bibitem{Talkner1}  P. Talkner, E. Lutz and P. H\"anggi,
 Phys. Rev. E {\bf 75} 050102(R)  (2007); P. Talkner, P. H\"anggi, J. Phys. A {\bf 40} F569 (2007). 
\bibitem{Gaspard1}   D. Andrieux and P. Gaspard, Phys. Rev. Lett. 
 {\bf 100}, 230404 (2008). 

\bibitem{SMukamel} M. Esposito, U.   Harbola and S. Mukamel, Rev. Mod. Phys  {\bf 81}, 1665 (2009). 
 
 \bibitem{Jar} C. Jarzynski,  D.K. Wojcik, Phys. Rev. Lett. {\bf 92}, 230602 (2004). 
 
\bibitem{Gaspard2} D. Andrieux,  P. Gaspard, T. Monnai
 and S. Tasaki, New J. Phys. {\bf 11} 043014 (2009). 
\bibitem{Campisi} M. Campisi, P. Talkner and P. H\"anggi,  Phys. Rev. Lett. {\bf 102}, 210401 (2009).

\bibitem{Lindblad} G.  Lindblad,  Commun. Math. Phys. {\bf 48}, 119 (1976).
\bibitem{Gardiner} C. W. Gardiner and P. Zoller,  {\it Quantum Noise},
 Springer (2000);  H.P. Breuer and F. Petruccione,
 {\it The theory of open quantum systems}, Oxford University Press (2002); 
 S. Haroche and J.-M. Raymond, {\it Exploring the Quantum}, Oxford Univ.Press (2006). 



\bibitem{Yukawa} S.  Yukawa, J. Phys. Soc. Jpn {\bf 69}, 2367 (2000).
\bibitem{CrooksQM}  G. E.  Crooks, J. Stat. Mech., P10023 (2008); 
  Phys. Rev. A  {\bf 77}, 034101 (2008).
\bibitem{Maes}  W. De Roeck and C. Maes,  Phys. Rev. E  {\bf 69}, 
  26115 (2004).

\bibitem{Roeck} W. De Roeck,  C. R. Physique {\bf 8}, 674 (2007).
\bibitem{SMukamel0} S. Mukamel,  Phys. Rev. Lett.  {\bf 90}, 170604  (2003).
\bibitem{SMukamel2}  M. Esposito, S. Mukamel, Phys. Rev. E {\bf 73}, 046129 (2006). 
 

 
\bibitem{Callen} H. B. Callen and T.~A. Welton, Phys. Rev. {\bf 83}, 34 (1951).
\bibitem{Kubo} R. Kubo, M. Toda and N. Hashitsume, 
  {\it Statistical Physics II: Nonequilibrium Statistical Physics} (1998). 

\bibitem{Hanggi}  P. H\"anggi and H. Thomas, Phys. Rep. {\bf 88} 207 (1982). 
 \bibitem{Agarwal} G . S. Agarwal, Z. Physik {\bf 258}, 409 (1972);
  W. A. Majewski, J. Math. Phys.  {\bf 25}, 3 (1984). 
 
\bibitem{Kurchan2} J. Kurchan,  J. Phys. A: Math. Gen. {\bf 31},
 3719 (1998); J.~L. Lebowitz and H. Spohn, J.  Stat. Phys. {\bf 95} 333, (1999). 
  

 
 \bibitem{Prost} J.  Prost, J.-F. Joanny, J.~M.~R.  Parrondo,  Phys. Rev. Lett.
 {\bf 103} 090601 (2009); R.  Chetrite, G.  Falkovich, K.  Gaw\c{e}dzki, J.Stat. Mech. P08005 (2008).
 

 \bibitem{Weidlich} W.  Weidlich, Z. Physik {\bf 248} 234 (1971).
 
 

\bibitem{Martin}  P. C. Martin and J. Schwinger , Phys. Rev. {\bf 115} 1342 (1959). 
 

\bibitem{Chetrite} R.  Chetrite and K.  Gaw\c{e}dzki,  Commun. Math. Phys.
 {\bf 282}, 469 (2008). 

 
\bibitem{Attal}  S. Attal, K. Gaw\c{e}dzki, Private communication.


\end{thebibliography}
\end{document}